# Forecasting Energy Needs with Logistics

Theodore Modis[*]

[*] Theodore Modis is the founder of Growth Dynamics, an organization specializing in strategic forecasting and management consulting: http://www.growth-dynamics.com
Address reprint requests to: Theodore Modis, Via Selva 8, 6900 Massagno, Lugano, Switzerland. Tel. 41-91-9212054, E-mail: tmodis@yahoo.com

**Highlights**

- By mid-21st century coal, oil, and natural gas will be the major energies comparable in size; hydroelectric will double.
- By mid-21st century renewables will have practically replaced nuclear energy but still at ¼ the size of the other energies.
- US oil production has followed a logistic trajectory extremely closely for 100 years.
- US oil production by fracking is 30% completed through its logistic trajectory.
- US oil production will represent less than 1% of the oil consumed worldwide by mid-21$^{st}$ century.

**Abstract**

The logistic function is used to forecast energy consumed worldwide and oil production in the U.S. The logistic substitution model is used to describe the energy mix since 1965 presenting a picture significantly different from the one covering the previous 100 years. In the new picture coal gently gains on oil and hydroelectric gains on natural gas even if it is three times smaller. Finally, renewables—wind, geothermal, solar, biomass, and waste—grow exclusively on the expense of nuclear, and are poised to overtake it by the late 2030s. By mid-21$^{st}$ century, coal, oil, and natural gas still remain the main players of comparable size. Hydroelectric has almost doubled in size. The only significant substitution is that of renewables having replaced nuclear albeit remaining at less than a ¼ the size of the other three energy sources. U.S. oil produced by fracking is forecasted to cease by mid-21$^{st}$ century, while oil produced by traditional methods should continue on its slowly declining trend. US oil production is likely to represent less than 1% of the oil consumed worldwide by mid-21$^{st}$ century.

# 1. Introduction

The logistic function is generally suitable to describe natural-growth processes. Originally designed for species populations it is dotted with predictive power as demonstrated by the fact that no ecological niche has ever remained incomplete under *natural* conditions. In nature, deviations from logistic trajectories occur whenever there are mutations, natural disasters, or other unnatural events. In society, deviations from logistic trajectories may result from "inappropriate" government decisions, wars, or major technological breakthroughs. Deviations meet with resistance and those resulting from ill-conceived decrees are generally short-lived permitting the logistic trajectory to continue its course toward completion.

     This work attempts to make reliable long-term forecasts for energy needs. The logistic function is used because it describes a natural law and therefore is dotted with a capability to make more accurate longer-term forecasts than simple curve-fitting techniques. The study concentrates on *physical variables* and shuns economic indicators. The latter, just like prices, are a rather frivolous means of assigning lasting value and will not follow a natural-growth process. Inflation and currency fluctuations due to speculation or politico-economic circumstances can have a large unpredictable effect on monetary indicators. Extreme swings have been observed. For example, Van Gogh died poor, although each of his paintings is worth a fortune today. The art he produced has not changed since his death; counted in dollars, however, it has increased tremendously. Therefore our analysis using logistics will not consider arguments intrinsically connected to prices such as energy return on investment (EROI).

     Fisher and Pry have demonstrated that the logistic function is also suitable to describe competitive substitutions.[1] It was in 1977 that Marchetti first used logistic substitution models to study the primary energy mix worldwide.[2] Software developed by Nakicenovic in IIASA implemented a generalized logistic-substitution process and permitted long-range forecasts for the world energy, which yielded an elegant picture that became classic.[3] The implication was that a few parameters that described well the trajectories of the entry and exit of primary energies on the world stage for more than one hundred years would continue doing so in the long-term future. Since that time there have been many publications and updates of this picture.[4][[5][6][7][8]

     As time went by persisting deviations from the model began making their appearance. They were generally dismissed as "to be soon reabsorbed" by proponents of logistics. Work coming out from IIASA maintained an unchanged model as late as 2002.[9][10] However, Fisher and Pry had warned us of substitutions that may not proceed to completion due to "locked" market segments.

     Devezas et al. formally addressed the important deviations from Marchetti's original energy-substitution picture. They wrote, "an astonishing deviation … can be observed … the pattern was broken and replaced by relative flatness from the mid-1980s."[11] In fact, upon more careful observation one can detect the beginning of such deviations as early as the mid-1960s. Devezas gives credit to Smil for first pointing out a significant deviation in the energy-substitution picture. In 2000 Smil proposed that Marchetti was wrong in concluding that the system's dynamics cannot be influenced, and stated that: ''After 1973, many forces began reshaping the system on a massive

scale…"[12] But it was Stewart who had pointed out a flattening of the energy trajectories more than ten years earlier.[13]

Arguments have been made for energy substitutions correlating with Kondratieff waves.[14][15] But there is evidence that these waves may now be deviating from their regular patterns.[16] Russian economists raise the possibility of even the end of this cyclical phenomenon altogether.[17] Could such evolutions have contributed to the energy deviations mentioned above?

Today it has become compelling to revisit those logistic forecasts of that elegant world-energy picture not only in order to check the validity of that model but also because long-term forecasts of energy needs always remains a topic of vital interest and discussion.

Oil production in the U.S. is a related subject and is treated in Section 5. The availability of extensive detailed and reliable data permits the evidence of a large, textbook-like, logistic fit, which depicts a recent deviation due to a "mutation" called shale oil. Shale oil production via hydraulic fracturing (fracking) follows a logistic trajectory of its own. The importance of shale oil cannot be overemphasized and there have been forecasts for its future ranging from one extreme to the other. The U.S. Energy Information Administration (EIA) predicts net exports to continue increasing through 2050.[18] But other voices forecast oil production by fracking to level off in the early 2020s.[19]

The analysis in this article assigns quantitative confidence levels on the uncertainties of forecasts whenever realistically feasible. The latest data available are used—up to May 2018—from the *BP Statistical Review of World Energy 2018* and from the EIA. Older data come from the *Statistical Abstract*, the Census Bureau of the U.S.

**2. Worldwide Energy Consumption**

There are historical data on world energy consumption going back to 1860.[20] In Figure 1 we see the yearly evolution of this variable up to the end of 2017. The data points generally follow a logistic trajectory albeit wiggling somewhat around it. In fact, one may want to discern a finer structure consisting of two or three smaller constituent logistics. However, it was decided to fit a single overall logistic to the entire data set (thick gray line) in order to obtain more reliable long-term forecasts. The data cover 70% of the logistic by the end of 2017. We can estimate the uncertainties on the parameters of the fitted logistic from look-up tables in the detailed Monte Carlo study by Debecker and Modis.[21] The uncertainty on the final ceiling of the logistic is estimated as ±14% and the midpoint in 1997 has an uncertainty of ±3 years, both with 90% confidence level. The dotted lines in the figure delimit this uncertainty on the forecasted trajectory. It means that nine times out of ten future values should fall inside the band delimited by the dotted lines.

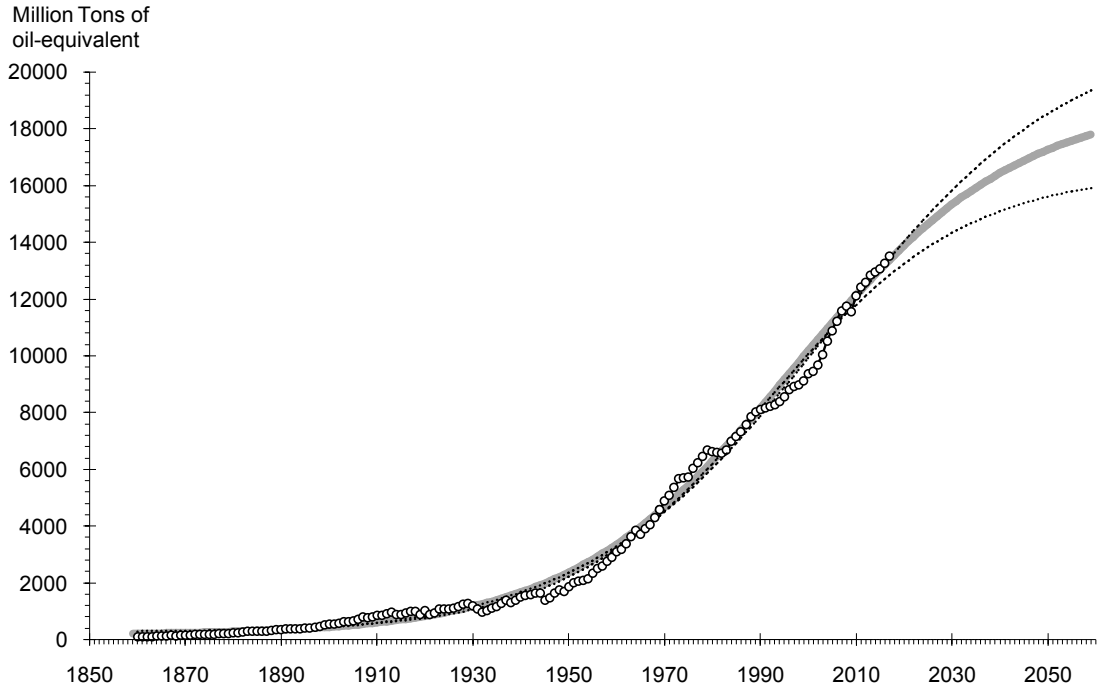

**Figure 1** Annual energy consumed worldwide (yearly data). The gray line is a logistic fit. The dotted lines delimit the 90% confidence level.

**Table 1: Forecasts and uncertainties for world energy consumption**

|  | Million tons of oil equivalent | Error |
|---|---|---|
| 2018 | 13386 | +1.9%<br>-3.0% |
| 2020 | 13719 | +2.3%<br>-3.5% |
| 2025 | 14498 | +3.2%<br>-4.6% |
| 2030 | 15196 | +4.2%<br>-5.7% |
| 2040 | 16345 | +6.1%<br>-7.7% |
| 2050 | 17194 | +7.7%<br>-9.2% |

### 3. The Primary Energy Mix

The classic graph on primary energy substitutions published by Marchetti in 1987 is reproduced in Figure 2 with the data updated to the end of 2017; the small circles designate updates. The graph shows the shares of primary energy sources plotted with a non-linear (logistic) vertical scale that transforms S-curves into straight lines.

It becomes immediately evident that some of the deviations from the smooth trajectories of the model, introduced as early as the mid-1960s—notably on the coal and the natural gas trajectories—not only did not become reabsorbed later but became more pronounced. The elegant model description that did fair justice describing the data for more than one hundred years is unquestionably no longer valid. From the four primary energies considered only oil behaved according to the model and even that shows persistent deviations during the last ten years.

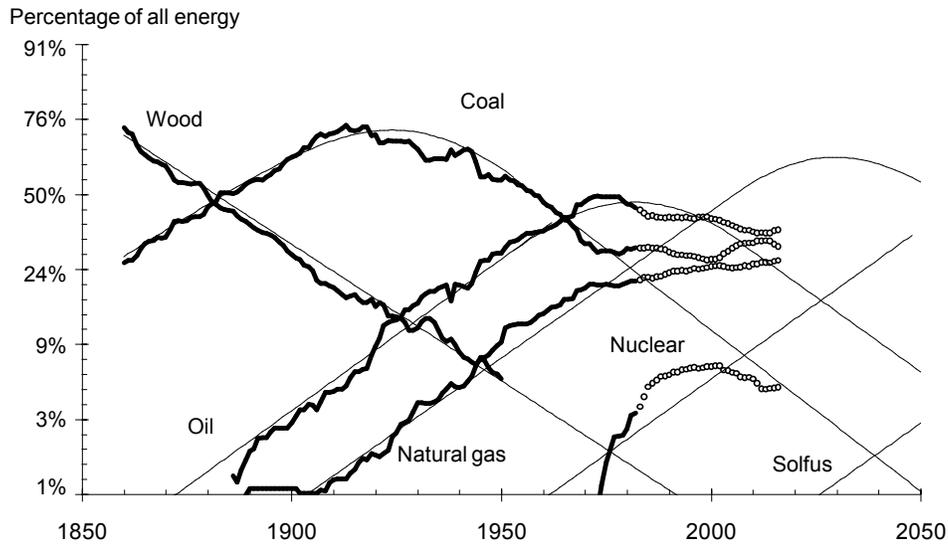

**Figure 2** Shares of primary energy sources. The black lines reproduce the picture published by Marchetti in 1987.[3] The small circles are updates up to the end of 2017. The thin black lines represent the logistic substitution model. The vertical scale is non-linear (*logistic*) and transforms S-curves into straight lines. "Solfus" is a hypothetical future energy source coined by Marchetti as a combination of solar and nuclear fusion.

At the same time, new renewable energy sources—comprising wind, geothermal, solar, biomass, and waste—have become progressively more important. Even the hydroelectric energy—neglected by Marchetti as insignificant (less than 1% share of the world market at the time)—can no longer be neglected. Moreover, there seem to be complementarities, such as between the coal and the oil trajectories. Therefore, a new attempt is undertaken below to use the logistic substitution model to describe an enhanced set of primary energy sources with combinations that will yield straight-line (logistic) behavior. The historical window begins in 1965—when deviations first appeared—and for which time period there are data from only one source, the *BP Energy Review*, which ensures data consistency.

### 4. Four Competitors

In Figure 3 coal and oil are grouped together, and so are natural gas and hydro resulting in a 4-competitor picture. The two subgroups, Coal + Oil and Natural gas + Hhydro, follow distinct straight lines implying a logistic behavior. The share of nuclear energy is saturating (i.e. it is calculated as 100% minus all other shares), and renewables enter steeply also along a straight line. For the sake of better continuity the thin straight lines of the model have been determined from the more recent trends of the data.

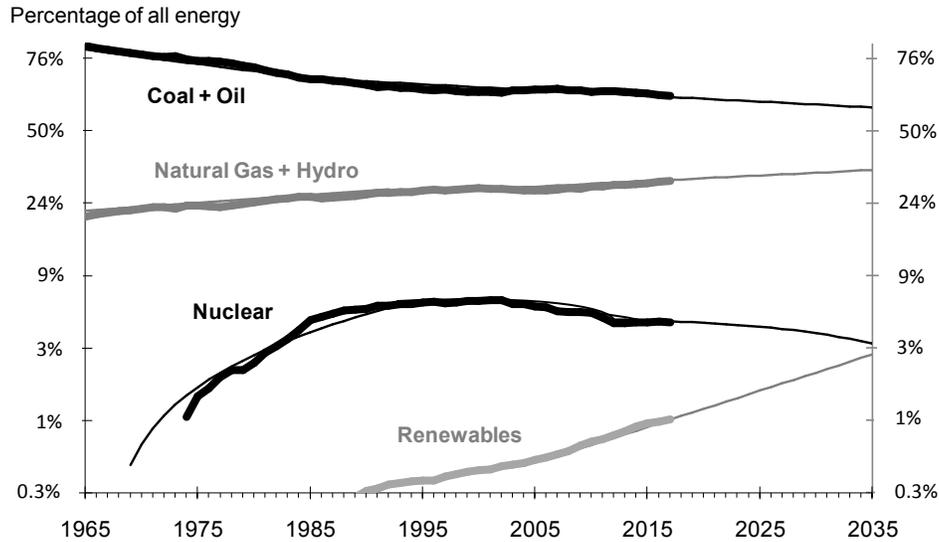

**Figure 3** Shares for the primary energy sources worldwide since 1965. The vertical scale is non-linear (*logistic*). The thick lines are annual data; the thin lines are descriptions by the logistic substitution model.

The forecasts for market shares from Figure 3 can be combined with the forecasts for the total energy from Table 1 to give forecasts in million tons of oil equivalent for each one of the four competitors.

**Table 2: Forecasts for the four competitors**

|  | Market Shares | | | | World totals | Energy forecasts | | | |
|---|---|---|---|---|---|---|---|---|---|
|  | Coal+Oil | Gas+Hydro | Nuclear | Renewables | Forecasts | Coal+Oil | Gas+Hydro | Nuclear | Renewables |
| 2018 | 63% | 32% | 4.6% | 1.1% | 13423 | 8434 | 4234 | 613 | 142 |
| 2020 | 62% | 32% | 4.5% | 1.2% | 13745 | 8577 | 4387 | 618 | 163 |
| 2025 | 61% | 33% | 4.2% | 1.6% | 14484 | 8880 | 4762 | 613 | 229 |
| 2030 | 60% | 34% | 3.8% | 2.1% | 15126 | 9108 | 5120 | 581 | 317 |
| 2040 | 58% | 36% | 2.5% | 3.7% | 16131 | 9354 | 5781 | 405 | 591 |
| 2050 | 56% | 38% | 1.0% | 5.4% | 16818 | 9373 | 6369 | 168 | 908 |

## 4. Two Microniches

*Coal + Oil*

In order to separate coal from oil we study the microniche Coal + Oil, see Figure 4. The two trajectories are obviously complementary but also rather flat. Straight-line descriptions (the logistic fits indicated with thin lines) show a practically stationary evolution with oil dominating and coal gently rising.

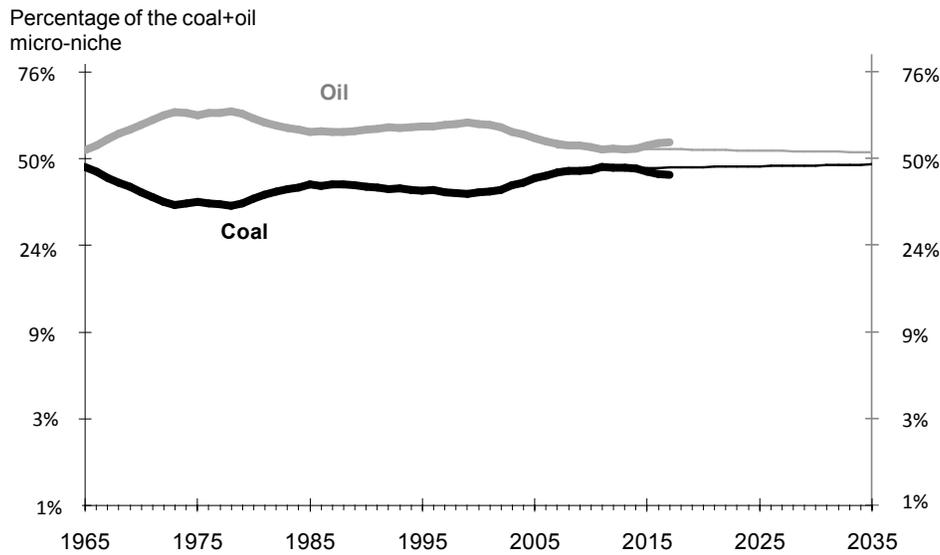

**Figure 4** Shares for oil and coal in the Oil + Coal microniche worldwide since 1965. The vertical scale is non-linear (*logistic*). The thick lines are annual data; the thin lines are descriptions by the logistic substitution model.

The forecasts for market shares from Figure 4 can be combined with the energy forecasts from Table 2 to give forecasts in million tons of oil equivalent for these two competitors.

**Table 3: Forecasts for oil and coal in million tons of oil equivalent**

|  | Marketniche shares | | | Energy forecasts | |
|---|---|---|---|---|---|
|  | Oil | Coal | Coal+Oil | Oil | Coal |
| 2018 | 53% | 47% | 8434 | 4474 | 3960 |
| 2020 | 53% | 47% | 8577 | 4535 | 4042 |
| 2025 | 53% | 47% | 8880 | 4670 | 4210 |
| 2030 | 52% | 48% | 9108 | 4764 | 4345 |
| 2040 | 52% | 49% | 9354 | 4839 | 4621 |
| 2050 | 51% | 49% | 9373 | 4794 | 4578 |

*Natural Gas + Hydroelectric*

In order to separate natural gas from hydroelectric we study the microniche Natural Gas + Hydroelectric, see Figure 5. The two trajectories are again complementary with natural gas three times as important. Straight-line descriptions (the logistic fits indicated with thin lines) show hydroelectric slowly gaining ground.

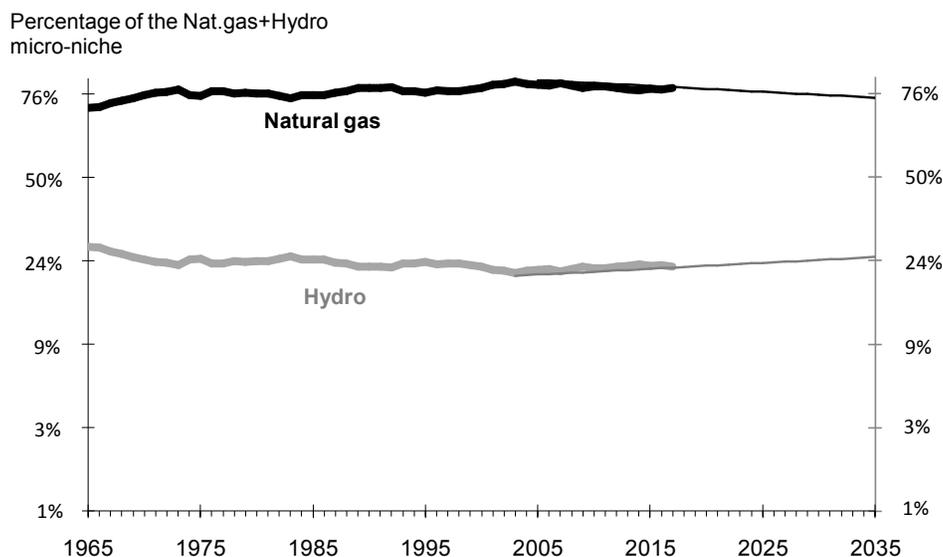

**Figure 5** Shares for natural gas and hydroelectric in the Natural gas + Hydro microniche worldwide since 1965. The vertical scale is non-linear (*logistic*). The thick lines are annual data; the thin lines are descriptions by the logistic substitution model.

The forecasts for market shares from Figure 5 can be combined with the energy forecasts from Table 2 to give forecasts in million tons of oil equivalent for these two competitors.

**Table 4: Forecasts for natural gas and hydro in million tons of oil equivalent**

|      | Marketniche shares | | | Energy forecasts | |
| --- | --- | --- | --- | --- | --- |
|      | Nat. gas | Hydro | Gas+hydro | Nat. gas | Hydro |
| 2018 | 78% | 22% | 4234 | 3150 | 907 |
| 2020 | 77% | 23% | 4387 | 3294 | 972 |
| 2025 | 76% | 24% | 4762 | 3522 | 1082 |
| 2030 | 76% | 24% | 5120 | 3733 | 1194 |
| 2040 | 74% | 26% | 5781 | 4030 | 1397 |
| 2050 | 73% | 27% | 6369 | 4399 | 1653 |

## 5. U.S. Oil: Two Kinds of Production

The U.S Energy Information Administration provides monthly data on U.S. oil production since 1920. Figure 6 shows that oil production closely followed a bell-shaped curve (the 1$^{st}$ derivative of a logistic) until 2008 or so, which corresponds to a level of 80% completion of the logistic curve. However, from 2008 onward there is a marked upward excursion due to shale oil produced via hydraulic fracturing; standard production of oil continued declining but production by fracking grew rapidly.

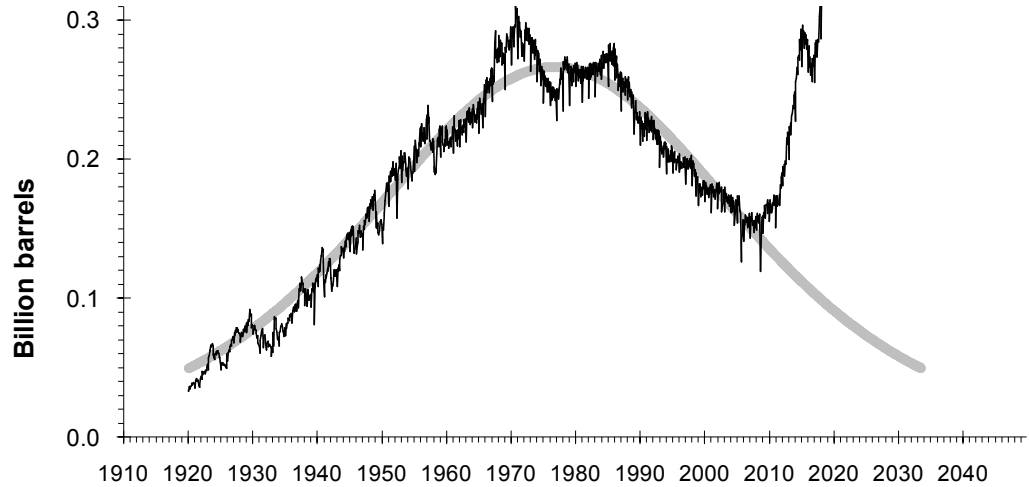

**Figure 6** US Crude Oil Production (monthly data). The pronounced peak on the right-hand side represents oil produced by fracking. The thick gray line is the 1$^{st}$ derivative of a logistic fitted on cumulative data up to 2008.

The EIA provides yearly data on the admixture of oil produced by fracking and oil produced by the standard method. In Figure 7 the substitution between these two processes takes on a logistic character. The trajectories of the two complementary shares are fairly-straight lines that cross at the end of 2015, beyond which point more oil is being produced by fracking than by standard extraction.

We can use the data from Figure 7—monthly numbers are obtained by interpolation—to separate total oil production into oil produced by the standard method and oil produced by fracking.

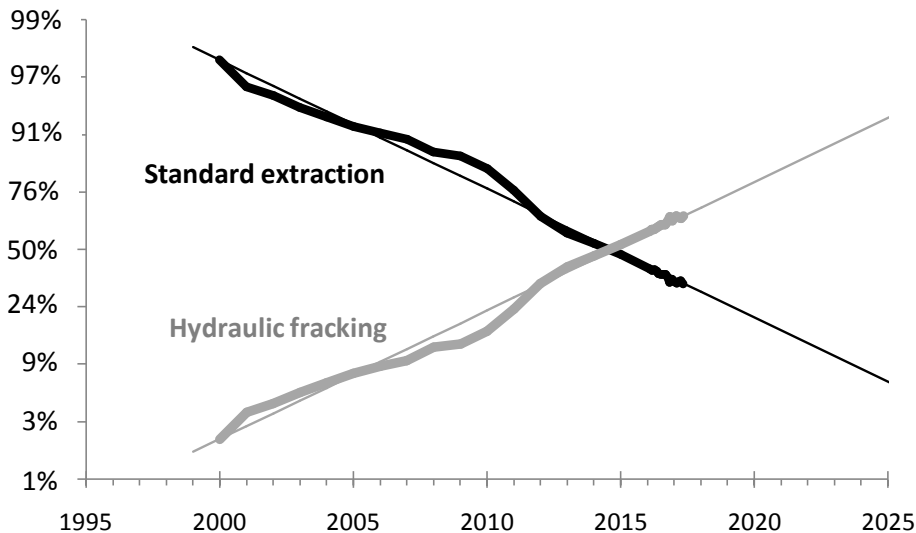

**Figure 7** The split between oil produced by fracking and oil produced by the standard method (data between 2000 and April 2018). The thick lines represent data; the thin lines the logistic substitution model.

*An Exemplary Logistic-Growth Process*

Figure 8 shows oil produced only by the traditional extracting method; Figure 9 shows these data cumulatively with a logistic fit—thick gray line—which is excellent. The estimated uncertainty on the level of the final ceiling—again from look-up tables in Reference [5]—is ±2.8% with 95% confidence level. The data cover 85% of the logistic and the midpoint is in 15-Jun-1976 with an uncertainty of ±3 months again with 95% confidence level.

In Figure 8 the monthly data trend fluctuates around the logistic life cycle. Notably there is a short peak between Jun-2012 and Mar-2017. But soon the production catches up again with the level of the declining logistic.

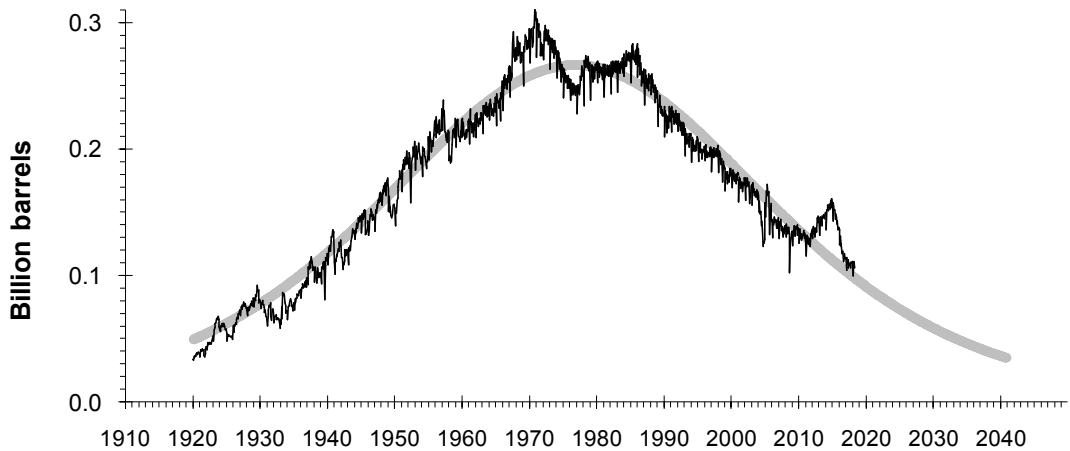

**Figure 8** US Crude Oil production by the standard method (monthly data). The thick gray line is the derivative of the logistic curve fitted on the cumulative data in Figure 9.

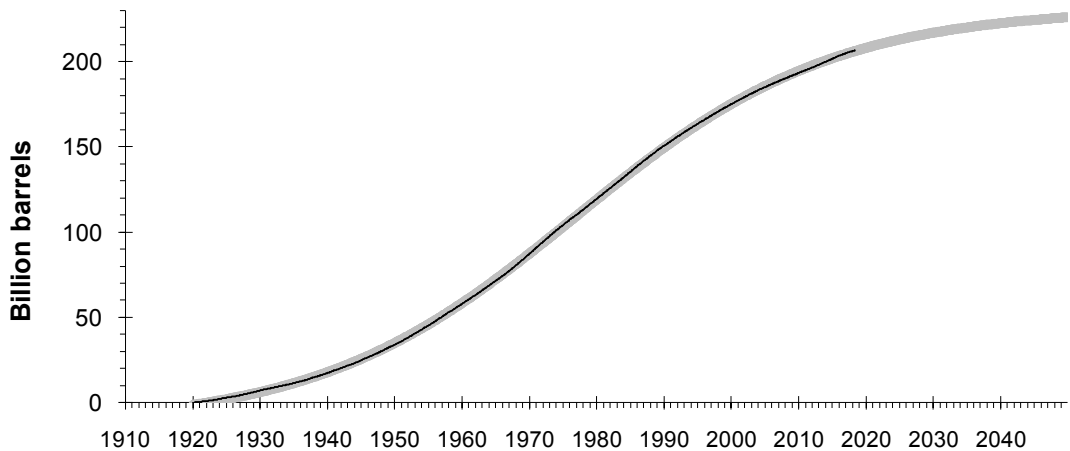

**Figure 9** US Crude Oil production by the standard method (cumulative monthly data); The thin black line represents the actual data. The thick gray line is a logistic curve fitted on the data.

*Oil Produced by Hydraulic Fracturing*

Figures 10 and 11 show data and logistic fit for oil produced by hydraulic fracturing, cumulative and monthly data respectively. Once again the fit is excellent but the uncertainties are now significantly larger. The estimated uncertainty on the level of the final ceiling is ±30% with 95% confidence level. The data cover 30% of the logistic and the midpoint is estimated in 28-Feb-2021 with uncertainty ±1.5 years. Table 5 shows the forecasts that arise from Figures 8 and 11.

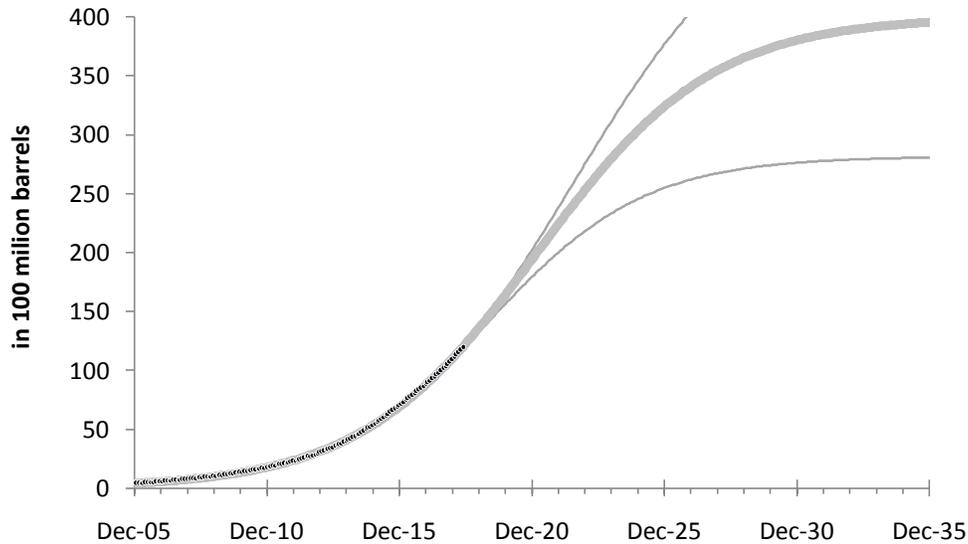

**Figure 10** U.S. Crude Oil produced by fracking (cumulative monthly data). The tiny black dots are the actual data points. The thick gray line is a logistic fit. The thin gray lines delimit the 95% confidence level.

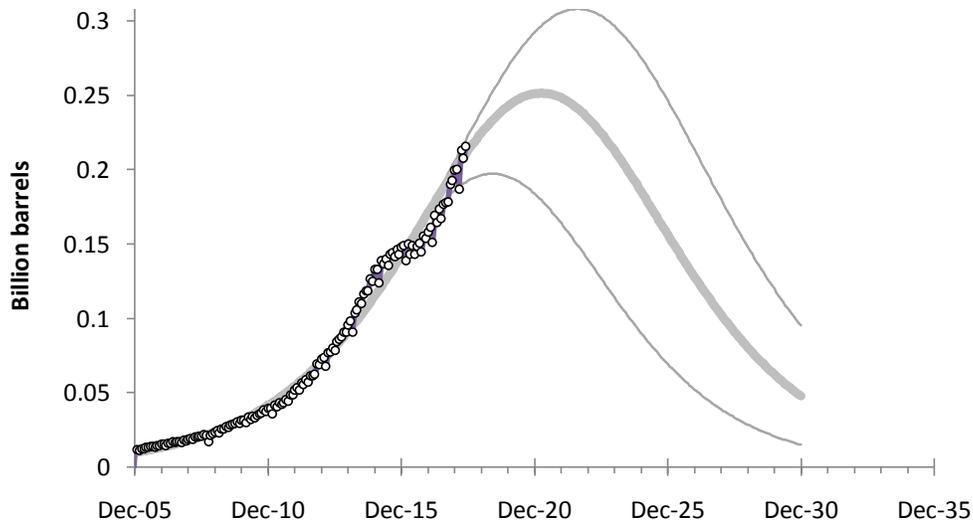

**Figure 11** U.S. Crude Oil produced by fracking (monthly data). The little circles are the actual data points. The thick gray line is the 1$^{st}$ derivative of the logistic curve in Figure 10. The thin gray lines delimit the 95% confidence level.

**Table 5: Forecasts for oil production in the U.S. in million barrels**

|      | Standard | Fracking | Total |
|------|----------|----------|-------|
| 2018 | 1700     | 2557     | 4257  |
| 2020 | 1075     | 2980     | 4055  |
| 2025 | 864      | 2037     | 2900  |
| 2030 | 688      | 651      | 1338  |
| 2040 | 427      | 35.5     | 463   |
| 2050 | 260      | 1.7      | 262   |

We can transform the Total numbers from Table 5 to million tons of oil equivalent in order to compare with the forecasts for worldwide oil consumption from Table 3. Table 6 below shows that the oil produced in the U.S. will represent less than 1% of the oil consumed worldwide by 2050.

**Table 6: Forecasts in million tons of oil equivalent**

|      | Oil produced in the US | Oil consumed wordwide | The US share |
|------|------------------------|-----------------------|--------------|
| 2018 | 596                    | 4474                  | 13%          |
| 2020 | 568                    | 4535                  | 13%          |
| 2025 | 406                    | 4670                  | 9%           |
| 2030 | 187                    | 4764                  | 4%           |
| 2040 | 65                     | 4839                  | 1.3%         |
| 2050 | 37                     | 4794                  | 0.8%         |

## 6. Discussion

World energy consumption has been described by a logistic curve spanning the last 150 years. The forecasts thus obtained indicate a slowing down of the rate of growth toward the mid-21$^{st}$ century. The historical window—1965 to 2018—was studied in detail in order to understand the deviations from previous energy forecasts. It was only after grouping the shares of oil with coal and natural gas with hydroelectric that it became possible to use the generalized logistic substitution model with success. This grouping into microniches does not only render the use of the substitution model possible; it makes broader sense. Coal and oil are both on the decline while natural gas and hydro are both on the rise. Moreover, environmentally speaking, coal and oil are more polluting fuels, while natural gas and hydro are the less polluting.

    Devezas addressed the same problem by grouping together oil with natural gas arguing that they have similar geological origins and locations, are extracted with similar technologies, often by the same commercial organization, and are often transported through pipes. His classification restores the validity of the substitution model but only up to the mid-1970s. From then onward he needs to take into account energy efficiency

calculations involving estimates for the world GNP. And he considers the specific future mix of renewables and nuclear energy as "uncertain."

In my opinion oil belongs with coal (while natural gas with hydro) because of their polluting nature and the growing environmental awareness. This is amply demonstrated in Figures 3 and 4 where less polluting energies are on the rise and more polluting on the decline. Devezas's arguments about similarities in extraction and transport techniques between oil and natural gas are of secondary importance. These arguments concern technical issues, as fracking was to standard oil extraction. Fracking may have altered the trajectory of U.S. oil production but it had a negligible effect on the trajectory of oil worldwide, whereas the shift toward cleaner energies is a fundamental phenomenon on a worldwide scale.

Moreover, my grouping validates the substitution model well into the $21^{st}$ century with no need to invoke efficiency arguments, which involve the unreliable monetary indicators. Energy efficiency has been steadily growing over the past decades (probably following a logistic curve of its own) and this has already been folded in in the calculation of the logistic forecasts for energy.

Finally, my grouping yields understanding of the detailed competitive dynamics. Within the two microniches, coal seems to be gently gaining on oil, while hydroelectric is gaining on natural gas even though it is three times smaller. And renewables—wind, geothermal, solar, biomass, and waste—are growing exclusively on the expense of nuclear and are poised to overtake it in mid-2035. All things considered, by mid-$21^{st}$ century, coal, oil, and natural gas are still the main players and of comparable sizes; hydroelectric has almost doubled in size. The only significant substitution that has taken place is that of renewables having practically completely replaced nuclear energy, albeit while remaining at less than a ¼ the size of the other three.

The uncertainties on the overall energy consumption forecasts—estimated via the look-up tables in Reference [21]—are not propagated down to the level of individual energy types because the forecasts on the shares carry additional uncertainties that cannot be reliably quantified.

The production of U.S. oil is considered separately as a species growing into its own niche. A logistic curve describes one hundred years worth of monthly data extremely accurately. However, in the $21^{st}$ century, oil extracted by hydraulic fracturing enters the picture as a "mutation", and introduces large deviations from the logistic trajectory justifying the consideration of the new extraction technique as a different species growing into its own niche. Oil extracted by fracking begins substituting oil extracted by the standard method along a logistic trajectory. Splitting overall oil production according to the two extracting methods we find two logistic growth processes: a long one spanning one hundred years that is 85% completed by the early 2018, and a short one spanning three decades that is 30% completed by the early 2018. It is interesting that when extraction by fracking picked up significantly, extraction by the traditional method also increased resulting in a small peak above the declining logistic trend for about five years. As Fisher and Pry have pointed out, when threatened by a new technology the incumbent technology redoubles its efforts in the competitive struggle. Nevertheless, a few years later the well-established declining logistic trend was regained. This phenomenon invites targeted research, namely to search for actions or events between Jun-2012 and Mar-2017 that caused extra production of oil being extracted with the traditional method.

Oil produced by fracking is forecasted to be a much shorter-lived process than oil produced by the traditional method. According to their logistic descriptions, which have been very successful up to now, fracking oil should cease production by mid-21st century despite its recent phenomenal success, while standard oil will continue being produced along its slowly declining trend. This conclusion may seem difficult to accept considering that besides the operational wells today there already exist many additional hydraulic fracturing wells awaiting exploitation, the so-called DUCs (Drilled but Uncompleted wells).[22] If the forecasts of Table 5 are to come true, the opposition by environmentalists may be responsible to quite an extent. The logistic function describes the law of *natural growth* and environmentalists are vehement advocates of natural growth, organic growth, and the like; they oppose fracking arguing that it damages nature.

As mentioned in the introduction natural-growth processes—described by the logistic function—proceed to completion *under natural conditions*. This should not be interpreted as the forecasts made in this article will come true if nothing is done. Natural conditions for a logistic fit include all the kinds of things that took place during the historical window used for the fit. Actions by environmentalists, technological breakthroughs, awareness of the public and the politicians, discovery of new reserves, and economic development worldwide have all been folded in while fitting the logistic. In the future policy makers will not have a free hand to do whatever they wish. They must continue listening to the growing voices of the public about climate change and the environment, and responding proportionally to the loudness of these voices, just as they have been doing in the past. The continuation of all well-established process on their *natural-growth* trajectories will ensure that the forecasts made in this paper come true.

Deviations could be caused only by the kind of major events never seen before tantamount to a "mutation" in the evolution of a species. But new breakthroughs in technologies such as artificial intelligence, bioengineering, or nanotechnology would not impact the evolution of most trajectories forecasted above because the beginning of these technologies has already been folded in the evolution of the data during recent years, and also we have seen this kind of technologies emerge regularly during the historical window. On the other hand, unseen-yet events of greater importance, e.g. major wars or the mastering of thermonuclear fusion would certainly interrupt/modify the forecasted trajectories.

Renewable are emerging as the rising stars of the future. An obvious suggestion for further study is the evolution of each one of them individually and the competitive dynamics between them.

**Endnote**

[1] Theodore Modis is a physicist, strategist, futurist, and international consultant. He is author/co-author to over one hundred articles in scientific and business journals and nine books. He has on occasion taught at Columbia University, the University of Geneva, at business schools INSEAD and IMD, and at the leadership school DUXX, in Monterrey, Mexico. He is the founder of Growth Dynamics, an organization specializing in strategic forecasting and management consulting: http://www.growth-dynamics.com